\input harvmac
\def \rr {({r_0 \over r})}

\def\a{\alpha}

\def \M {{\cal M}}
\def \F {{\cal F}}
\def \N {{\cal N}}
\def \rf { \refs}
\def \te {\textstyle }

\def \diag {{\rm diag}}
\def \f {{\rm f}}
\def \hN {\hat N} 

\def \V {{\cal V}}
\def \s{{\rm t}} 
\def \k {\kappa} 
\def \F {{\cal F}}
\def \g {\gamma}
\def \del {\partial}

\def \ha{{\textstyle{1\over 2}}}

\def \a {\alpha}
\def \b {\beta}
\def \ov {\over}
 
\def\r {\rho}
\def \p {\phi}
\def \m {\mu}
\def \n {\nu}

\def \l {\lambda}
\def \t {\theta}
\def \td {\tilde }
\def \d {\delta}

\def \gym { g^2_{\rm YM} } 
\def \l {\lambda}

\def \inv {^{-1}}
\def \ov {\over }
\def \four{{\textstyle{1\over 4}}}

\def \d {\delta}

\def \lr { \lref}
\def \k {\kappa}

\def \M {{M}}

\def\np {  Nucl. Phys. }
\def \pl { { Phys. Lett.} }

\def \pr  {  Phys. Rev. }

\lr \bbpt {K. Becker, M.  Becker, J.  Polchinski  and  A.A. Tseytlin,
Higher order graviton scattering in M(atrix) theory,
 Phys.  Rev. D56 (1997) 3174,   hep-th/9706072.
 } 
\lr\cht{ I. Chepelev and A.A. Tseytlin,
Long distance interactions of branes: Correspondence between supergravity and superYang-Mills
descriptions,  
 Nucl. Phys.  B515 (1998) 73, hep-th/9709087.  
}

\lref\mald{J. Maldacena, Probing near-extremal black holes with D-branes,  Phys. Rev. D57  (1998) 3736,  hep-th/9705053.}
\lref\malda{J. Maldacena, Branes probing  black holes,
Nucl. Phys. Proc. Suppl. 68 (1998) 17,  hep-th/9709099.
}
\lref\kapus{J.I. Kapusta,  Finite-temperature field theory
(Cambridge Univ. Press,  1989). }
\lr\frt{E.S. Fradkin and A.A. Tseytlin,
Quantum properties of higher dimensional and dimensionally reduced supersymmetric theories, Nucl. Phys. B227 (1983) 252. }
\lr\gkp{S.S.  Gubser, I.R.  Klebanov
  and A.  Peet,
Entropy and temperature of black 3-branes,
Phys. Rev.  D54 (1996) 3915, 
 hep-th/9602135.  }
\lr\gkt{S.S. Gubser, I.R.  Klebanov and A.A. Tseytlin, 
Coupling constant dependence in the thermodynamics of \N=4 supersymmetric Yang-Mills theory, 
Nucl. Phys. B534 (1998) 202, 
hep-th/9805156.}
\lr\TT{A.A. Tseytlin, 
Open superstring partition function in constant gauge field background at finite temperature, Nucl. Phys.  B524 (1998) 41, 
hep-th/9802133.}

\lr\pat{
J. Pawelczyk and S. Theisen,
$ AdS_5 \times  S^5$ Black Hole Metric at $O(\alpha'^3)$,
 hep-th/9808126.}

\lr \KT {I.R. Klebanov and A.A. Tseytlin, 
Entropy of near-extremal black p-branes,
\np { B475} (1996) 164, 
 hep-th/9604089.}

\lr\matt{T. Banks, W. Fischler, S.H. Shenker and L. Susskind, 
M theory as a matrix model: a conjecture, Phys. Rev. D55 (1997) 5112, hep-th/9610043.}

\lr \duff{
M.J. Duff  and J.X. Lu, 
The selfdual  type IIB  superthreebrane,
{Phys. Lett.}  {B273} (1991)  409. }

\lr  \hs { 
G.T. Horowitz and A.~Strominger, Black strings and p-branes,  Nucl.
  Phys. B360 (1991) 197. }

\lr\osl{X.  Kong   and F.  Ravndal,
Quantum Corrections to the QED Vacuum Energy, 
Nucl. Phys. B526 (1998) 627, 
hep-ph/9803216.} 
 
\lr\DPS { M. Douglas, J.  Polchinski and  A.  Strominger, 
Probing five-dimensional black holes with D-branes, 
 JHEP 12  (1997) 003,
 hep-th/9703031. }
\lr\polch { J. Polchinski, Dirichlet Branes and Ramond-Ramond charges, 
Phys. Rev.
  Lett.   75  (1995) 4724,
 hep-th/9510017.
}

\lr \dkps{  M.R. Douglas, D. Kabat, P. Pouliot and S.H. Shenker,
 \np B485 (1997) 85,
hep-th/9608024. } 

\lr \dougl{  M.R. Douglas and  M. Li, 
D-brane realization of $N=2$ 
 super Yang-Mills theory  in four dimensions,
hep-th/9604041;
M.R. Douglas and W. Taylor, Branes in the bulk of 
Anti de Sitter space, hep-th/9807225.
} 

\lr \MM { J.~Maldacena, The Large N limit of superconformal field theories and
  supergravity,
{Adv. Theor. Math. Phys.} {2} (1998) 231,
  hep-th/9711200.}

\lr \lin{D.A. Kirzhnits and A.D. Linde, \pl B42 (1972) 471; 
S. Weinberg, \pr D9 (1974) 2257; L. Dolan and R. Jackiw, 
\pr D9 (1974) 3320.}

\lr \kleb{ 
I.R. Klebanov, World volume approach to absorption by nondilatonic branes,
  {Nucl. Phys.} {B496} (1997) 231,
  hep-th/9702076.}
\lr\GKT{
S.S. Gubser, I.R. Klebanov and A.A. Tseytlin, String theory and classical
   absorption by three-branes, Nucl. Phys. B499 (1997) 217,
 hep-th/9703040. } 

\lr \DS {
 M.  Dine   and N.  Seiberg,  Comments  on higher derivative operators
in some susy field theories, 
    Phys. Lett. B409  (1997) 239,  
    hep-th/9705057;
S.  Paban, S. Sethi  and M.  Stern, Supersymmetry and higher derivative
 terms in  the effective action of Yang-Mills theories, 
        JHEP 06  (1998) 012, 
    hep-th/9806028.  }

\lr\GKP{ 
 S.S. Gubser, I.R. Klebanov   and  A.M. Polyakov,
 Gauge theory correlators from noncritical string theory, 
 Phys. Lett. B428 (1998) 105, 
 hep-th/9802109.}

\lr\WW{E. Witten, Anti-de Sitter space and holography, 
Adv. Theor. Math. Phys. 2 (1998) 253, 
 hep-th/9802150.
}

\lr \miao {M. Li, Evidence for large $N$ phase transition
in $\N=4$ super Yang-Mills theory at finite temperature,
hep-th/9807196.}

\lr \bala {V. Balasubramanian, P. Kraus, A. Lawrence and S.P. Trivedi, 
Holographic probes of Anti de Sitter spacetimes, 
hep-th/9808017.}

\lr\kapu{J. Kapusta, D. Reiss and S. Rudas, 
Transition from weak to strong coupling in QCD and in grand unified models at high
temperature, 
\np B263 (1986) 207. }

\lr\bgr{
T. Banks and M.B.  Green, 
Non-perturbative effects in $AdS_5 \times S^5$ string 
theory and $d=4$ susy Yang-Mills,
J.High Energy Phys.05 (1998) 002, 
  hep-th/9804170.}

\lr\horr{
G.T. Horowitz and  S.F. Ross,
Possible resolution of black hole singularities from large $N$ gauge theory,
JHEP 04 (1998) 015,   hep-th/9803085.}

\lr\wi{
E. Witten, Anti de Sitter  space, thermal phase transition, 
and confinement in gauge theories, 
Adv. Theor. Math. Phys.  2 (1998) 505, 
 hep-th/9803131.} 

\lr\soo{
 S.-J.  Rey, S. Theisen  and  J.-T. Yee,
Wilson-Polyakov loop at finite temperature in large $N$ gauge theory, 
Nucl. Phys. B527 (1998) 171,  
 hep-th/9803135.}

\lr\bran{
 A. Brandhuber, N. Itzhaki, J. Sonnenschein  and S. Yankielowicz,
Wilson loops in the large $N$ limit at finite temperature, 
hep-th/9803137.}

\lr \peet{A.W. Peet and J. Polchinski, 
UV/IR relations in AdS dynamics, hep-th/9809022.}

\lr\stev{S.S. Gubser and I.R. Klebanov, Absorption by Branes
 and Schwinger Terms in the World Volume Theory,  Phys. Lett. B413 (1997) 41, 
hep-th/9708005.}


\baselineskip8pt
\Title{\vbox
{\baselineskip 6pt
{\hbox {Imperial/TP/97-98/072    }}{\hbox{TAUP-2517-98  }}
{\hbox{hep-th/9809032}} 
{\hbox{   }}
}}
{\vbox{\vskip -30 true pt
\medskip
\centerline {Free energy  of $\N=4$ super Yang-Mills in Higgs phase}
\medskip
 \centerline {and     non-extremal D3-brane interactions }
\vskip4pt }}
\vskip -20 true pt 

\centerline  {A.A. Tseytlin$^{1,2}$\footnote{$^{\star}$}
{\baselineskip8pt e-mail address:
tseytlin@ic.ac.uk}\footnote{$^{\dagger}$}{\baselineskip8pt
Also at  Lebedev  Physics
Institute, Moscow.} 
and  S. Yankielowicz$^3$\footnote {$^*$} 
{e-mail address: shimonya@post.tau.ac.il   } 
}



\medskip

\medskip
\smallskip\smallskip
\centerline {$^1$\it  Theoretical Physics Group, Blackett Laboratory}
\smallskip
\centerline {\it  Imperial College,  London SW7 2BZ, U.K.}
\medskip
\centerline{$^2$\it  Department of Physics and Astronomy }
\smallskip
\centerline{\it  Rutgers University,  Piscataway NJ 08854-8019, USA  }
\medskip
\centerline {$^3$\it   School of Physics and Astronomy}
\smallskip 
\centerline {\it  Beverly and Raymond-Sackler Faculty of Exact Sciences}
\smallskip
\centerline {\it 
Tel Aviv University, Ramat Aviv, 69978,
 Israel
 }

\bigskip\bigskip
\centerline {\bf Abstract}
\baselineskip10pt
\noindent
\medskip
We study the free energy of  $\N=4$ super Yang-Mills theory
 in the  Higgs phase 
 with a mass scale $ M $ corresponding to non-zero v.e.v.
of the scalar fields. At zero temperature this theory 
describes  a system of parallel
separated extremal D3-branes. Non-zero temperature
corresponds to non-extremality in supergravity description.
 We interpret  the supergravity interaction
potential between a non-extremal D3-brane and a D3-brane probe
 as  contribution of massive states 
to the free energy of the $N\to \infty$ SYM theory at strong  't Hooft 
coupling ($\l = 2N \gym \gg 1$). Both  low ($M \gg T$) and high
($T \gg M$) temperature regimes are considered. For  low temperature 
 we find that the structure of terms 
that appear in the free energy  at strong and weak coupling is the same.
 The analysis
of the high-temperature regime  depends on a  careful identification 
of the scalar field v.e.v. in terms of the 
 distance between branes  in the supergravity description 
  and  again  predicts  strong coupling terms  
similar to those found in the weak-coupling $\N=4$ SYM theory.
 We 
consider also the  corrections to the strong coupling results 
by taking
into account the leading $\a'^3 R^4$ string contribution to the 
supergravity  effective action. This gives
rise to the  $\l^{-3/2}$  corrections 
 in the coefficient functions of  $\l$
 which
multiply  different 
 terms in  free energy.
                            
\bigskip
 
\Date {September 1998}


\noblackbox \baselineskip 15pt plus 2pt minus 2pt 

\newsec{Introduction}
D-branes,   having both  closed string and open string descriptions
 \rf{\polch,\dkps}  
provide an important link between certain  physical quantities  in 
supergravity and super Yang-Mills theory (see, in particular, 
\refs{\gkp,\kleb,\GKT,\DPS,\MM, \GKP,\WW}).
While  $N$ extremal  D3-branes 
are described by a vacuum state of $\N=4$ SYM theory,  
the  near-extremal D3-branes  are  expected to be described 
by a  thermal  state of the SYM theory.

The interpretation  of near-extremal supergravity  D-brane 
 solutions as  corresponding to finite temperature states on SYM side 
is very natural in view of the existence of the Hawking temperature 
and entropy associated with them.
 Indeed, the Bekenstein-Hawking entropy
of  non-extremal D3-brane has the same form as the  entropy
of the conformal SYM theory \gkp. The relative
 coefficient of $3/4$ between the weakly coupled SYM result
and the B-H entropy \gkp\  reflects the need  to interpolate from  strong to weak coupling in 
order to compare the  supergravity result to  the 
perturbative  SYM expression for the entropy \gkt.

Here we would like to address the question of 
a similar 
 relation  between  the SYM thermodynamics in the  massive  Higgs   phase 
(or Coulomb branch corresponding to the case of non-zero vacuum expectation values of
the  scalar fields)  and  the  supergravity description of  non-extremal 
 D3-branes. As in \gkt\  the 
 aim will be to  interpret the supergravity results 
as predictions  about SYM free energy in the strong coupling regime.

Having non-zero temperature $T$  and non-zero  scalar expectation values $U \sim M$ 
in SYM theory should correspond to  a system of  
separated non-extremal D3-branes.  We shall consider 
the simplest possible case when $N\gg 1$ coinciding D3-branes  (`source')
 are separated  
by a distance (proportional to a mass scale  $ M $) 
from a single-charge  D3-brane (`probe').  
We shall  assume that  the parallel 
branes are separated along the  
single transverse direction, i.e. the $U(N+1)$
scalar (coordinate)  matrix  is $\hat U_n = \diag( \delta_{n1} U, 0, ...,0)$, 
breaking the gauge symmetry $U(N+1) \to U(N) \times U(1)$. 
We shall be interested in the 
case when the  whole system is 
 in a thermal state with temperature  $T$  proportional to the 
non-extremality parameter of the D3-brane solution.
The aim will  be to compute the supergravity 
interaction potential\foot{Some previous related discussions appeared in
\refs{\DPS,\mald,\malda,\TT}.}
  and to interpret it as a 
contribution of massive states 
to the free energy of the $N\to \infty$ 
 SYM theory  at strong 't Hooft  coupling ($\l = 2 N \gym \gg 1$).

At zero temperature  the static  interaction potential between 
the source and the probe vanishes -- parallel  branes  represent  a 
supersymmetric BPS  state.  At finite temperature 
it is natural to expect that there will be an attractive
potential between the branes, i.e. there will be a tendency
to restore the broken gauge symmetry.

This  expectation is in agreement  with the  fact that 
the extremal D3-brane solution \refs{\hs,\duff}  with separated centers does {\it not}
 have a non-extremal generalisation which is static --
the only  static 
non-extremal solution is a single-center black  
D3-brane \hs.  In general, non-extremal  black holes 
attract  and fall onto each other,  i.e. one obtains a static solution only if they are at the same point
(we exclude the possibility of having an infinite periodic array of black holes).
Extremal   BPS black holes  or branes 
form a much larger space of  static solutions  parametrised by
  harmonic functions, i.e.   they can be separated and put at different points
at  no const  in energy.

This is what  happens 
in  gauge  theory at weak coupling:  if we  start with a vacuum state
with a non-zero scalar value $U \sim M $ and  switch on the temperature
 $T \gg M$,
the  induced scalar potential $ a_1T^4  + a_2 T^2 U^2 + ... $
 will drive the system back to the  unbroken phase with $<U>=0$ \lin. 
The finite temperature breaks supersymmetry and lifts the vacuum degeneracy
of $\N=4$ SYM theory.
 This is seen at weak coupling, and 
we interpret the above remark about 
supergravity solutions
as the  prediction that the same happens 
 also at  large $N$ and  strong 't Hooft coupling.

Below we shall consider  the supergravity expression for the 
interaction potential between non-extremal
D3-brane `source' and D3-brane `probe'. We  shall 
interpret it in
terms of the
massive states contributions to the SYM free energy  $F$ as
 a function of the temperature $T$ and mass scale $M\sim U$ (or separation between branes)
in Higgs phase at strong coupling, and  also compare  it  with the
weak-coupling  SYM theory  result for the  free energy.
We shall find that   while the  weak and strong coupling expressions 
contain terms of similar structure, 
their coefficients are,  in general,   different.  This 
follows the same  pattern as discovered  in \refs{\gkp,\gkt}
for the coefficient in  the    free energy $\sim V_3 T^4$  in the massless
(coinciding brane) case.  As in \refs{\bgr,\gkt}\ we shall consider the modification 
of the supergravity  expressions by the leading $\a'^3 R^4$ 
string correction  and  suggest  (and  present evidence) 
 that at finite 't Hooft coupling  $\l$
different terms in $F(T,M)$ are  multiplied  by 
 non-trivial functions of $\l$, which at strong 
coupling go as $ a + b \l^{-3/2}+ ...$.\foot{We are assuming
that there is no phase transition at finite $\l$
(for some arguments to the opposite 
see \miao).}

We shall mostly concentrate on  the case   when  separation is large compared to non-extremality, 
i.e.  when the temperature is much smaller than the mass scale.
When $M \gg T$ the contribution of the  massive states to  the SYM 
free energy  we are interested in  may be computed in two steps.
One may first ignore the temperature  and  integrate  (in the $N\to \infty$ limit) 
the massive modes out
obtaining a   Wilsonian effective action  ($W \sim \int d^4x \sum \l^{l-1} {F^{2l+2}\ov M^{4l}}$)
for the massless modes.
One may  then switch on the temperature and compute the
free energy  starting with this effective action.
On the supergravity side, 
the probe's action  in an extremal source background  is 
known  to be related to $W$.  Given that  the structure of the static potential  
part of the probe's action  in a non-extremal  background is very similar
to that of the  velocity-dependent part  in the extremal case,
it may  not be  completely  surprising that one finds  also 
a close  correspondence   between  the   thermal average of $W$ and 
the interaction potential of a D3-brane probe with a non-extremal  D3-brane source.

In section 2  we shall  compute  the supergravity interaction potential 
 as a function of non-extremality parameter ($r_0 \sim T$) 
and separation ($r \sim M$).
We shall  consider  the  large  charge 
 (or `near-core')  approximation.

In section 3 we shall  study   the `low-temperature' or 
`long-distance' ($M \gg T$) expansion of the resulting free energy 
and compare the first two leading terms 
($c_1T^4 + c_2 {T^8\ov M^4}$) 
with  the corresponding terms in the 
weak-coupling SYM free energy. The ${T^8\ov M^4}$
term will be shown to a have a 3-loop origin in  the SYM perturbative 
expansion. In general, the  ${T^{4l+4}\ov M^{4l}}$ term  in $F$ 
has  $(l+2)$-loop origin in the SYM free energy  and is 
related to 
the term $F^{2l+2}\ov M^{4l}$ in the effective Wilsonian action
at $T=0$.

In section 4  we shall  discuss  the opposite 
`high-temperature' or 
`short-distance' ($T \gg M$) expansion. 
The comparison of this expansion with the corresponding 
SYM expansion  {\it a priori} may seem  less reliable. Unlike the previous case
where the strings between the source and probe are long and,
therefore, represent very massive states, in the present case the
temperature fluctuations are much larger then the Higgs scale (or 
separation between branes). Nevertheless, in view of the successful understanding 
of the thermodynamics of coinciding branes, it would be interesting to
see how much of the physics in this regime is accounted for by the 
supergravity expansion. 
As explained  in  Appendix,  while 
the $M$-dependent part of the 1-loop free energy vanishes exponentially in the 
$ M \gg T$ limit,   expanded for $T \gg M$ it contains
 the well-known  $T^2 M^2 $, \ $T M^3$, etc.,  terms.
Similar terms are found in the  supergravity potential, 
 suggesting that they should be 
present  in the strong-coupling SYM free energy.

In section 5  we consider the $\l^{-3/2}$  corrections 
to the coefficients in the  strong-coupling expression  for  the free energy
produced by taking into account the leading $\a'^3 R^4$ string  contribution
to the supergravity action.

In Appendix  we present   the expression for the 1-loop 
effective potential in $\N=4$ SYM theory at finite temperature.

\newsec{ Supergravity expression for
D3-brane interaction potential}
We shall consider 
 a D3-brane  probe (with charge 1) in  a  non-extremal D3-brane  background 
(with charge $N$) and  find 
 the static interaction  potential as a function of non-extremality parameter
$r_0$ and separation $r$.  This potential  will be attractive and will
vanish at large separation (it will vanish also 
  for $r_0\to 0$ corresponding to the  BPS limit). 
It 
 will  be  interpreted
as a free energy of a $U(N+1)$  SYM  gas  at  finite temperature (related to $r_0$)
and  non-zero scalar vacuum expectation value  or mass scale  $M$ (related  to $r$).

The
metric of  the non-extremal D3-brane  and the `electric' components 
of the 4-form potential  are ($i=1,2,3$) 
\eqn\one{
ds^2 = H^{-1/2} (- \f dt^2 + dy_i dy_i)
   +  
   H^{1/2} ( \f^{-1}  dr^2 + r^2 d\Omega_5^2) \ ,  
}
$$ C_{0123} =  H^{-1} (1-H_0) \ , $$
where
\eqn\hhh{
 \ H_0 =1 + {L^4 \ov r^4} \ ,  \ \ \ \ \ \ H=1 +  {\td L^4\ov r^4} \ , \ \ \ 
 \ \ \ \
 \f=1 -  {r^4_0\ov r^4 } \ , 
}
\eqn\lll{
L^4 = 4 \pi g_s N \alpha'^2 \ , \ \ \ \ \ \ \ \ \  
\td L^4 =  \sqrt{ L^8 + \four r^8_0} - \ha r^4_0
\ .   }
The string coupling $g_s$ is related to the  SYM  coupling 
$g_{\rm YM}$  and the  't Hooft coupling $\l$ by
\eqn\coup{
\gym =  2 \pi g_s \ , \ \ \ \ \  \ \ \ \ \ 
\l\equiv    2 N \gym = 4\pi N g_s = L^4 \a'^{-2} \ . }
The non-extremality parameter is related to the Hawking temperature $T$
 as  follows  \rf{\gkp,\KT}
\eqn\temp{
r_0 = \pi L^2 T \ . }
We  shall consider D3-brane  probe to be static and  parallel  to
the  source 
and choose the  static gauge (that such configuration is 
not a classical solution without external force  will not be important for us). 
We shall assume that the euclidean time (same for the probe and the source)
has  period  $\b = { 1 \ov T}$. 
Then the interaction part of the action for the D3-brane 
 probe  takes the form\foot{Similar  actions were discussed in  \refs{\DPS,\mald, \malda}.}
\eqn\acti{
I = T_3 V_3 \b    H^{-1} \big(  \sqrt \f  -1 + H_0 -  H\big) \ , 
}
where $V_3$ is the volume of the spatial directions along the D3-brane
and $T_3$ is the D3-brane   tension \polch
\eqn\ten{
T_3= { 1 \over 2\pi g_s (2\pi \alpha')^2}  = { N \ov 2 \pi^2 L^4} \ . 
}
We shall assume that 
$N \to \infty, \  \l \gg 1$ for the validity of the
 supergravity description \refs{\kleb,\GKT,\DPS}
and also 
consider the low-energy  or  gauge  theory 
 limit \refs{\kleb,\malda,\MM}
in which 
\eqn\limi{
\a' \to 0\ , \ \ \ \ \ \ \ \  \bigg\{
U={ r\ov \a'}\  , \ \ U_0 = { r_0\ov \a'} = \pi \l^{1/2} T\  ,\  
\ g^2_{\rm YM}= 2\pi g_s \ , \ \ T \ \bigg\} \   = \ {\rm fixed} \  .   }
Equivalently, one may set $\a'=1$ but take  
$L \gg r, \ \ L \gg r_0$ (i.e.  $\td L \to L$).   In this limit  
\eqn\lii{
 H \to  H_0 \to  { L^4 \ov r^4} =  { \l \ov U^4} \ , \ \ \ \ \ \  \ \ \   \f 
= 1 - {r^4_0 \ov r^4} =  1 - { U_0^4
 \ov U^4} \ .  }
The resulting metric \MM\  that  follows from  \one\   has horizon at $U=U_0$ 
and a curvature singularity at $U=0$ \horr. The finite temperature  $\N=4$ SYM  theory
 should  describe the  full geometry including the under-horizon region \horr.\foot{The 
gauge theory in 
a thermal state should  represent the 
D3-brane in thermal equilibrium  with a gas  of string modes in $AdS_5$ \refs{\MM,\horr}.} 
The vielbein components of the  Weyl tensor    are proportional 
to $ U^4_0 \ov \l^{1/2} U^4$ and  are small  near (and outside) the horizon
where  one can thus  trust the supergravity description. 

The action of  the   probe  placed  outside  of  the horizon 
 \acti\  now  becomes
\eqn\onee{
I \equiv \b F =   \b V_3  T_3 L^{-4} r^4 \bigg[ \sqrt{ 1 -
 {r^4_0\ov r^4}}  -1\bigg] \  . 
}
The corresponding    `free energy' is 
\eqn\iiu{
F =  V_3  { N  U^4  \over 2 \pi^2  \l^2 }  \ \bigg[ \sqrt { 1 -  
   {\pi^4  \l^2 T^4 \over U^4 }  } - 1 \bigg] \ , 
}
or, equivalently, 
\eqn\iiuu{
F = { \te { 1 \ov 2 \pi^2}}   V_3  N  \M^4   \ \bigg[ \sqrt { 1 -  
   {\pi^4 T^4 \over \M^4 }  } - 1 \bigg] \ , \ \ \ \ \ \ 
\ \ \  \M\equiv {r \ov  L^2} = { U \ov  \l^{1/2} }   \ . 
}
$\M={r \ov  L^2}$ and $T = {r_0 \ov \pi L^2}  $
  are the natural  parameters  corresponding to 
    the supergravity solution  with the characteristic  scale $L$.
However,   in the  comparison  with the SYM perturbation theory 
in section 3.2 the role of the mass parameter will be played by $U$.
Note that the temperature mass scales  in the  SYM  theory 
are $M_0=T$ (tree-level fermion  mass),  
$M_{el}= \l^{1/2} T$ (1-loop scalar mass and the `electric' gauge boson mass) and $M_{magn} = \l T$ (magnetic screening scale at which  finite $T$  theory becomes non-perturbative). 
$M_{el}$  is associated with    the non-analytic $ M^3_{el}T =\l^{3/2}T^4$ term
 in the free energy  \kapu.

Below we  will 
 interpret  the  expression  \iiu\ 
as the $N\to \infty, \ \l \gg 1$ limit 
of the 
SYM free energy at temperature $T$  in 
the phase with  a  non-zero scalar expectation 
 value   $U$.

The supergravity potential  \iiu\ if interpreted as 
the exact expression for the SYM  free energy in the strong-coupling
limit   suggests the existence of the  maximal temperature
(for given $U$ and $\l$). The square root formula \iiu\
has the same `Born-Infeld' origin  (cf.\acti) 
as the action of a relativistic 
D-particle, so the bound 
on  $T$ is similar  to the maximal velocity or maximal 
field strength bounds.  
Since \iiu\ is formally valid in the low-temperature phase, 
this suggests that 
$T_c = \pi\inv \l^{-1/2} U$
may be a  `critical temperature'  at which a phase transition takes 
place.\foot{While there is no phase transition 
in the conformal case ($U=0$) \wi, it may  be present in the Higgs phase.}
Note that the entropy and  heat capacity  corresponding to 
the free energy \iiu\ 
\eqn\een{
 S =  - { \del F \ov \del T }  
=    {  N \pi^2 V_3  T^3 \over  (1 -   { \pi^4 \l^2 T^4 \over U^4 })^{1/2} } 
\ ,  \ \ \ \ \   C= T { \del S \ov \del T }
=  { 3N \pi^2 V_3  T^2  (1  -    {\pi^4  \l^2 T^4 \over 3 U^4 })
 \over  ( 1 -     {\pi^4  \l^2 T^4 \over U^4 })^{3/2} } \ 
}
diverge at the critical temperature.

In the above expressions we did not  yet make any assumptions  about 
the  relative values of $r$ and $r_0$ (or $U$ and $U_0$). 
As in \refs{\mald,\bran,\soo} we shall think of 
 the non-extremal D3-brane as  `located' 
at the horizon.  Thus the limit of large separation between the branes
or `small temperature' $M \gg T$ limit  is $r \gg r_0$ (or $U \gg U_0$) 
while the limit of  small separation  or `high temperature' $T \gg M$ 
is $r \simeq  r_0$ (or $U \simeq U_0$).

\newsec{ Long-distance  or   low  temperature  expansion}

Let  us first consider the long-distance 
approximation $r \gg r_0$ corresponding to the 
probe  being  far   away from the source. Equivalently, let us  
assume that  the temperature is very low, 
i.e. the scalar expectation value or the mass parameter
$U$ is  much larger than $U_0=\pi \l^{1/2} T$. 

\subsec{Leading term}
 The leading term in  the long-distance  or  low-temperature expansion of 
\iiu\ is 
\eqn\yey{ 
F= - \ha  T_3 V_3    L^{-4} r^4_0
  =   - \four N  \pi^2 V_3 T^4   \  . 
}
Remarkably,  this expression  is perfectly consistent  with  the 
 previous results   
\rf{\gkp,\gkt} about the correspondence  between the  entropy 
 or free energy of the massless
SYM gas  and  the  Bekenstein-Hawking entropy 
of near-extremal D3-brane.  The  free energy of $U(\hN)$
SYM theory in the  zero scalar (massless)  vacuum
is given (for $\hN \gg 1$)  by \gkt
 \eqn\fef{
F =   - {\textstyle { 1 \ov 6} } f (\l) \pi^2 V_3  \hN^2   T^4  \ , \ }
\eqn\faf{ 
f (\l \to 0)\ \to \ 1 \ , \ \ \ \ \ \ \ \ \ \ 
 f (\l\gg 1)  = {\te { 3 \ov 4} }  + { \te {45 \ov 32} } \zeta (3) \l^{-3/2} + ... \ 
\ \to \ \    {\te { 3 \ov 4} }  \ .    } 
Taking $$\hN = N +1$$ one thus   finds that  in 
 the strong coupling (supergravity) limit
\eqn\nnn{
F= 
 -   {\textstyle {1 \ov 8} }  (N^2  + 2 N + 1)   \pi^2 V_3 T^4   \ . 
}
In the weakly coupled SYM theory 
the separation of one of the D3-branes  from  $N$ others 
corresponds to switching on  a non-zero scalar 
 expectation value.\foot{Since $U^2 = U_n U_n $ \  
($n=1,...,6)$  we may assume that 
branes are separated along $n=1$ direction,  i.e. take  the $U(N+1)$
scalar matrix as
$ U_1 = \diag( U, 0, ...,0)$.}
Then the $O(N^2)$ and $O(1)$ terms in the total free energy  \nnn\ correspond
 to  the contributions of the  modes  which remain massless
while  the   $O(N)$ term  is the contribution of the massive
modes. 
The  $O(N)$ term  in \nnn\ is  exactly the expression in \yey\
 which  
represents  the 
energy of the separated brane.
 This is  consistent with the  interpretation  (see, e.g.,  \dougl) 
of the  probe's action  as being  a  result of intergrating out  the 
massive SYM modes (or strings connecting probe and source):
the analogue of 
\yey\  at weak coupling 
is the contribution to the total free energy produced by the loop of  massive 
modes.

  In general,  one  could  expect  that the  total   large $N$ 
free energy of the SYM gas in the  non-zero scalar (or `one brane separated') 
phase may 
have the form 
 \eqn\feif{
F =   - {\textstyle { 1 \ov 6} }  \pi^2 V_3  T^4 \bigg[ f  (\l) N^2 + 2\td  f (\l) N  + ...\bigg]
    \ .  \ }
What we have found is that  the functions $f$ and $\td f$ have the same 
weak and strong coupling limits and thus are likely to coincide also 
for finite  't Hooft coupling.  
We shall present a non-trivial check of this conjecture  in section 6
by showing that $f$ and $\td f$ do have the same $O(\l^{-3/2})$ terms in their
strong-coupling expansions.

\subsec{Subleading term}
The first   subleading term in the  large-distance expansion of \iiu\
 is 
\eqn\tyte{
F_1
=   -  {\te { 1 \ov 16}} 
 V_3  \pi^6  N   \l^2    { T^8  \ov U^4}     \  .  }
The supergravity predicts that this term  should be present 
in the strong-coupling limit  of the  SYM theory  free energy
in the `massive'  phase.
As we shall argue below, like in the $T^4$  term 
case, there is a similar  $T^8 \ov U^4$ term also in the weak-coupling expression
 for the SYM free energy. This term   has a  3-loop origin.

It is useful  to draw analogy  between the large-distance 
expansion of the non-extremal ($T\not=0$)  static potential \onee,\iiu\ 
and the
 velocity-dependent  D3-brane interaction 
 potential  in the extremal case\foot{Note that as in \iiuu\
one can rescale  the coordinates $\rm r$=$r\ov L^2$, 
 $\rm v$=$v\ov L^2$, absorbing  all powers of $L$, so that \ 
$\V  = { \te { 1 \ov 2 \pi^2} } N  V_3 
{\rm  r}^4 \bigg[ \sqrt { 1 -  { {\rm v}^2\over{\rm  r}^4 }  } - 1 \bigg] \ . 
$}
\eqn\ty{
 \V  = T_3 V_3  L^{-4} r^4 \bigg[ \sqrt { 1 -  { L^4 v^2\over r^4 }  } - 1 \bigg] 
    \  . 
}
The leading $v^2$  term corresponds to the  classical 
$U(1)$ \ $F^2_{\m\n} $  term on the SYM side,  the  $v^4\ov r^4$ term 
is reproduced by the  1-loop $F^4\ov r^4$   term, 
 the $v^6\ov r^8$ term -- by the 2-loop $F^6\ov r^8$ 
term, etc. (see, e.g.,  \rf{\dkps,\matt,\bbpt,\cht} and refs. there). 
The reason for the agreement between the  supergravity expressions 
and the weak-coupling  SYM  results  lies in the  known \DS\ and expected 
non-renormalisation  theorems
 of the relevant   terms in the  large $N$ 
SYM effective action:  the  perturbative $l$-loop  terms 
$(N\gym)^{l-1} { F^{2l+2}\ov M^{4l}} $  
 should not be multiplied by  functions 
of $\l=2 N \gym$ 
 \refs{\bbpt,\stev,\cht}.

At the same time, 
no such non-renormalisation theorems are expected  in the non-extremal or  finite  temperature case 
where supersymmetry is broken.   As already 
found on the example of the $T^4$ term \gkt,
different terms  in the free  energy may be multiplied 
by non-trivial  functions of the  SYM coupling, i.e. 
one may  find  the following expression 
for the  `massive'  contribution to the free energy 
\eqn\tyyt{
F (U \gg T)  =  - {\textstyle { 1 \ov 3} }  \pi^2 V_3 N   T^4  \bigg[ f(\l)  
+  h_1 (\l)  \l^2  { T^4 \over U^4 } + h_2 (\l)  \l^4  { T^8 \over U^8 }
 + ...  \bigg]  \ , }
 where,  like $f$ in \faf,  the functions $h_n$ approach 
finite values  both  at small and large $\l$,  i.e. 
\eqn\guu{
h_n(\l\to 0) \to a_n\ , \ \ \ \ \ \ \ \ \
h_n (\l \gg 1) \to  b_{n} +  c_n\l^{-3/2} + ... \ . }
The coefficients $b_n$ are determined  by 
the supergravity action \iiu. 

Let us now argue that   \tyyt,\guu\ is consistent also with 
the weakly coupled  SYM theory.
In general, the perturbative (planar)   loop expansion 
for the  $\N=4$ SYM free energy  as a function of temperature 
and scalar  vacuum value  or effective mass parameter 
$ M=U$   has the following  form (determined in view of  the 
UV finiteness of the theory   by 
 dimensional considerations)\foot{The  field   
normalizations in  the  SYM Lagrangian are 
$ L \sim { 1 \ov \gym} \tr ( F^2 +  D  U  D  U  +  U^4 + ...) $.}
\eqn\free{
F =  V_3 T^4 \bigg[ \F_1 ({ T \ov U})  + \l \F_2 ({ T \ov U})  + 
 \l^2 \F_3 ({ T \ov U}) + ... \bigg] \ . 
}
To compare with long-distance 
expansion in supergravity we are interested in the  low-temperature limit when 
  $U  \gg T$.
Then  it is easy to show (see Appendix) 
 that the  non-constant 
$U$-dependent part of the  1-loop correction
  vanishes  exponentially 
in this limit, 
 $\F_1 - { \pi^2 \ov 6}  \to  \ a\ {U^2 \ov T^2}  e^{-\sqrt \pi {U\ov T}} \to 0$. 
The same conclusion happens to   apply also   to $\F_2$.
However, the {\it 3-loop}  term $\F_3$ 
does contain  a nontrivial contribution
proportional to $ T^4\ov U^4$ which thus has the same structure as 
the strong-coupling (supergravity) correction \tyte. This is 
in agreement  with  the ansatz  \tyyt,\guu.

A  simple way to argue 
 that  the 3-loop  term   in  the SYM free energy 
contains a non-trivial $T^8\ov U^4$ contribution   is  the following.
The 1-loop SYM effective action at $T=0$ is known to  have  the 
$N {F^4\ov U^4}$ term (its supersymmetric extension) 
in the low-energy, large $U$ expansion.
Here $F$  is  the $U(N)$  massless gauge field strength
remaining after integrating out massive  string modes ($F^4$ 
stands for several terms with different  Lorentz index contractions). 
The thermal averaging   of $N {F^4\ov U^4}$ to 
the leading order in gauge coupling
then   leads to the 
required term (see also  \osl\ for analogous discussion 
in the context of QED). 
Indeed, 
\eqn\thrr{
 < F^4 > \ = \ \bigg[ c (N\gym)^2   + O((\gym)^4) \bigg] T^8  \ . }
The resulting  coupling dependence   is  as appropriate for a 3-loop
 planar  graph  contribution to   the free energy (averaging is done
with $e^{- F^2/\gym}$ measure). The $T^8$ behaviour 
follows on dimensional grounds.\foot{To  demonstrate that 
the expectation value  at $T\not=0$  is proportional to  $T^8$ 
it   is enough to consider  the abelian approximation.
Separating first  the points in $F^4$ as, e.g., 
 $ < F^2(x) F^2(0) > $  
  we get   terms like  $\del \del G \del \del G, $   where
 $ G $ is the massless  scalar propagator at finite $T$,
i.e. $G (x) \sim    \sum_n  [\ |x|^2 + (x_0- nT\inv)^2\ ]^{-1}$.  
Differentiating and  taking 
the limit of coinciding points 
we obtain  the $T^8$ result (using that the UV cutoff dependence 
 should cancel out as it is absent at $T=0$).}

Remarkably, the supergravity (strong-coupling) result in \tyte\ 
$r^8_0\ov r^4$ $\sim$ $T^8\ov U^4$  also has  the coefficient  $(N\gym)^2$.
Though we did not attempt  compute the  proportionality constant in the weak-coupling 3-loop 
SYM expression \thrr, there is no reason to expect that it will
coincide with the  corresponding constant 
 in the supergravity expression \tyte:
the weak-coupling and strong-coupling 
coefficients $a_1$ and $b_1$ in \guu\ are most likely different 
(as they are in \faf).  We shall confirm the presence of the $\l^{-3/2}$
correction in the strong-coupling expansion of the function $h_1 (\l)$
in \guu\ in  section 5, suggesting,  by analogy with  the $T^4$ term \gkt, 
that there is, indeed, a non-trivial 
renormalisation of the coefficient of the $\l^2 {T^8\ov U^4} $ term
in the free energy.

Similar conclusions seem to apply to other  terms in the 
low-temperature expansion of the free  energy \tyyt: the structure of
the terms $T^4 ( \l^2 {T^4\ov U^4})^l$  is the same  in the 
 weak and strong coupling
limits  while their coefficients may be different. 
Again, to argue that   such terms appear  (at $2l +1$ loop order)
in the weak-coupling expression
for free energy one may start  with 
the  fact that there is the $\l^{l-1}  {F^{2l+2}\ov U^{4l}}$  term 
in the $l$-loop  zero-temperature low-energy SYM effective action
(which, as was mentioned above, should 
 reproduce the $v^{2l+2}\ov r^{4l}$ term in \ty).
Taking the expectation value of the  $F^{2l+2}$ operator 
  at $T\not=0$ and in the planar limit 
 (that adds $l+1$  to the total loop order)
 one gets the required factor
 of $\l^{l+1} T^{4l+4}$.  Note that the 
 presence of the non-trivial functions $h_n(\l)$ in \tyyt\
(i.e. a breakdown of non-renormalisation theorems)
may be attributed to the second step of thermal averaging (cf.\thrr).

\newsec{Short distance   or high temperature  expansion}

The perturbative  free energy of  the  SYM  theory with mass  scale $M$
has   the following    {\it high} temperature
or small   mass ($ T \gg M$)  expansion (see Appendix):
\eqn\fre{ F( T \gg M)\   = \  a_1 T^4 + a_2 T^2 M^2 +  a_3 T M^3 + ... \ , }
 i.e. the functions  $\F_n$ in  
 \free\   ($M\sim U$) are \ 
$
\F_n( T \gg M)  = k_n    +  p_n { M^2 \ov T^2} +  q_n {M^3 \ov T^3 } + ... $.
To see if the  supergravity expression \onee\
representing strong-coupling limit of the SYM free energy 
contains similar $ T^2 M^2$,  etc.,   terms 
one is to consider  its short-distance or near-horizon $r\to r_0$ 
(or $U\to U_0$) 
expansion.
The action \onee\  will be describing a D3-brane 
probe separated by a small distance from the  horizon $U=U_0$ 
of  a 
non-extremal D3-brane source.\foot{One may wonder if the limit of small separation
between branes should correspond to the  expansion near $r=0$
where the brane source is supposed to be located in the extremal case. 
Note, however, that in the limit \limi\  or   $L\gg r,r_0$ 
both $r$ and $r_0$ are effectively sent to zero. 
In any case, the supergravity
description  is valid  for $U\geq U_0$ but is not valid
for $U\to 0$.}

One immediate  question   is how to relate the supergravity coordinate 
 $ r$  in the region  where it is close to  $r_0$ 
to the  scalar condensate or  mass scale $M$
of SYM theory.  
The relation between the  scalar  field   value   $U\sim M $
and $r$  here appears to be  more complicated  than the direct proportionality 
$r \sim   M $ which  applies 
for 
large $r$ 
  (cf.   \limi).

The  required
identification is  suggested  by the  proper choice of coordinates in the 
near-horizon region. 
As  was  noted in \mald,  to compare the  supergravity  description with  the 
SYM  theory one in the non-extremal case 
 one should  relate  the SYM scalars not  to  $r$ but  to  the 
new  `isotropic' coordinate $\rho$  defined by 
\eqn\coop{
 \f^{-1} dr^2  + r^2 d\Omega_5^2 =  \chi (\rho) (d\rho^2 + \rho^2 d\Omega_5^2)  \ ,     }
\eqn\huh{
\rho =  \exp \int {  dr \ov   r \sqrt \f} \   
=    \bigg( r^2 + \sqrt{r^4 -r^4_0} \bigg)^{1/2}\ ,  \ \ \ \ \ \ \ 
  r^2 = { \r^4 + r^4_0 \ov 2 \r^2 } \ .  }
Note that  $ \r \to \r_0=r_0$  when   $r \to r_0$. 
Expressed in terms of $\rho$ 
the free energy in  \onee\ becomes
\eqn\uuu{
 F = - \ha  V_3  T_3 L^{-4} r^4_0 \bigg( 1 +  { r^4_0 \ov \r^4} \bigg)   \  .
}
This change of coordinates  was  not esential 
in  the  above 
 discussion of the  large distance expansion,\foot{Note that 
with the choice of normalisation of $\r$ we made 
for large $r$  we get $ \r  \to \sqrt 2 r$. Then 
 \uuu\ is exactly the same as  the sum of the  two leading terms 
\yey,\tyte\  in the large $r$ expansion of \onee.}
but is  crucial  for  the  correct  SYM interpretation
of the probe's action in the  near-horizon  region. 

Expanding  $\r$ near $r_0$  and  introducing  the scalar mass parameter 
  $M$ as  (see \coup,\temp; cf.\iiuu) 
\eqn\nam{ M = {\r -r_0 \ov L^2} \ , \ \ \ \ \ \ \ 
\r -r_0  = \a'  \l^{1/2} M  \ ,  \ \ \ \ \  {\rm i.e.} \ \ \ \ \ \ \
\r = \a' \pi \l^{1/2} (T +  \pi^{-1}  M) \  ,  
}
we  find\foot{The inclusion of the $  \l^{1/2}$ factor in  the definition of $M$
in \nam\  may seem  optional
but as in \iiuu\ it leads to the  strong-coupling expression
for the free energy which does not 
include explicit powers of 't Hooft coupling. 
Similar identification of the energy scale
(as the scale associated with supergravity probes)
 was recently  suggested  in  \peet. 
If we absorb the  $ \l^{1/2}$    factor in $M$ then 
the high-temperature expansion below 
will contain extra $\l^{-n/2}$ factors.}
\eqn\eee{
F =  - \four \pi^2 V_3 { N  T^4   } \bigg[
 1 +  { 1 \ov   (1   + { M \ov \pi  T})^4 }  \bigg] \ .  } 
For small $M$ or high $T$ (by  `high temperature' expansion
we  mean  a near-horizon  near-extremal expansion where 
both  $M $ and $T $  are small but $ T \gg M$) we get
\eqn\exx{
{ F}  
= - \four \pi^2  N   V_3  \bigg(  2 {   T^4  }     
-   {    4\pi^{-1}   T^3 M   } 
+   {  10\pi^{-2}  T^2 M^2  } 
-   {  {20 \pi^{-3} } T M^3   }  + ... \bigg)  \ . } 
Let us discuss the meaning of the  several leading  terms in \exx. 
 The first  $T^4$ term  in \eee\  (whose coefficient 
 does not  actually 
depend on how we relate $r$ to $M$ at $r\to r_0$)
is similar to the $r\to \infty$ result \yey\
but  its coefficient is  different by the  factor of 2. 
It  has  the following  SYM interpretation. 
For small $r \sim r_0$  we have  a $U(N+1)$ SYM theory  describing all $N+1$ \ 
D3-branes close to each other.
 Taking  large $r \gg r_0$  corresponds to separating 
 one brane  so that we are left with a massless  $
U(N)$ theory.  It is the difference
$
F(r\to r_0) \ - \  F (r\gg r_0)   
$
that  should have the interpretation 
of the SYM   free energy, i.e which should 
account for the change in the number of relevant degrees
  of freedom. 
 Indeed,  this difference gives us
$  - \four \pi^2 V_3   N  T^4$ which happens to coincide 
 with $F (r\gg r_0)$ and   
 is exactly
in agreement with  order $N$ term in the (strong-coupling)
SYM result  \nnn.

The   $  T^2 M^2 $  and $TM^3$ terms are  the same 
as the ones that appear  in   the standard  
high-temperature expansion of the perturbative  (e.g., 1-loop) 
free energy of a massive theory (see (A.11)).
However, the $T^3 M$ term present in \exx\ is absent in the weak-coupling free energy.  
In general, it is natural to suggest   that
the  large $N$, finite $\l$  free energy has the following
high-temperature expansion  (cf.\tyyt)
 \eqn\eyt{
F (T \gg M)  =  - {\textstyle { 1 \ov 3} }  \pi^2 V_3 N   T^4  \bigg[ 2 f(\l)  
+  \k_1 (\l)   { M \over T } + \k_2 (\l)  { M^2 \over  T^2 }
  + \k_3 (\l)  { M^3 \over  T^3 }
+ ...  \bigg]  \ , }
where 
\eqn\finc{
 \k_n (\l\to 0 )\  \to\    w_n\ , \ \ \ \ \ \
\ \ \
\k_n (\l\gg 1  )\  \to   \  s_n + v_n \l^{-3/2} + ... 
\ .   } 
The  weak and strong coupling  coefficients  $w_n$ and $s_n$  
are,  in general,  different.
In particular,  $w_1=0$ but $ s_1\not=0$,  while  both 
$w_n$ and $s_n$ are  non-vanishing for $n>1$.

\newsec{$\a'^3 R^4$  string correction and   
  strong-coupling expansion of  SYM  free energy}
To justify our expectation  about the presence of $\l^{-3/2}$ corrections 
to  the strong-coupling expressions for 
the SYM free energy \feif,\tyyt,\eyt\ let  
 us now repeat the supergravity computation of section 2 now taking the
non-extremal D3-brane  source metric to be 
deformed by the leading $\a'^3 R^4$
correction  to the string effective action (see \refs{\bgr,\gkt} and refs. there). 
As in \gkt\ the idea  is  to  find  the 
 changes in the probe's  action  and thus in the free energy. This
will allow us to determine the
leading  correction to the strong coupling  SYM result as 
predicted by  string-corrected supergravity. 
Remarkably, we will  find that  the 
`$N^2 + 2N +1$' pattern   for   the coefficient of the 
leading  $T^4$
term in the action  \nnn\  survives the correction, implying  that  
the order $N$  (`massive')  contribution   should be  multiplied by the same 
function \faf\ of the  't Hooft coupling as found in the 
massless phase in \gkt.

Let us first  summarize the result for the corrected  
near-horizon (or large $N$, large $N g_s$)  limit 10-d Einstein-frame 
metric 
\refs{\gkt,\pat}:
\eqn\mett{
  ds^2_{10 E}  = e^{-{ 10 \over 3}\nu (x)}g_{5mn} (x) dx^m dx^n
 + e^{2 \nu(x)} L^2 d\Omega_5^2
\ . }
Here  the 5-d metric is \gkt\  (cf.\one)
\eqn\fif{
g_{5mn} (x) dx^m dx^n
= L^{-2}  r^2 (  - \ \f\ \f_1  dt^2 + dy_i dy_i)
   +  \f^{-1} \f_2  L^2 r^{-2} dr^2  \ , 
}
$$
\f = 1- \rr^4 \ , \ \ \ \ \ \ \
 \f_1 = 1 - 15 \g \bigg[ 5 \rr^4 + 5 \rr^8  - 3\rr^{12}\bigg]  + O(\g^2) \ , 
$$
\eqn\few{
\f_2 =1  +  15 \g \bigg[ 5 \rr^4 + 5 \rr^8  - 19\rr^{12}\bigg]  + O(\g^2)
\ , } 
and  the `$S^5$ raduis' function is \pat
 \eqn\raa{
e^\nu =1 +  {15\over 32}\g  \bigg[\rr^8 + \rr^{12}\bigg]
 + O(\g^2)
\ . } 
Note that $\nu$ does not have $ \rr^4$ term  as this function 
corresponds  to  a massive (fixed) scalar \gkt.  For completeness, 
the expression for the 10-d dilaton  which is needed
to determine the 10-d string-frame metric 
$ds^2_{10} = e^{- {1 \ov 2}  \p} ds^2_{10E}$
is  \gkt\
\eqn\dil{
e^{\p} = 1 - { 15 \ov 16} \g \bigg[ 6 \rr^4 + 3 \rr^8 + 2 \rr^{12}\bigg] 
+ O(\g^2)
\ . }
Here 
$\g$ is proportional to the coefficient of the $R^4$ term in the type II
superstring 
effective action  (compared to  \gkt\  we make $\g$ 
dimensionless by multiplying it by $L^{-6}$,  cf.\coup)
\eqn\gam{
 \g \equiv  { 1\ov 8} \zeta(3)\alpha'^3 L^{-6}
 = { 1\ov 8} \zeta(3) \l^{-3/2} \ . }
To find the static potential term in  the  action of a 
D3-brane probe  propagating in this 
background we also need the `electric' part of the 4-form potential.
The 
 corresponding self-dual 5-form field strength
  is determined by  the 
5-part of the above 10-d  Einstein-frame metric \mett\ 
(note that $F^2_5$ term is Weyl-invariant in 10 dimensions):
\eqn\reew{
F_{0123r} = \del_r C_{0123} = 4 L^{-1}   \sqrt{- h_5} \ , 
\ \ \ \ \  \ \ \  h_{5mn} \equiv   e^{-{ 10 \over 3}\nu }g_{5mn} 
\ . } 
Since  $ \sqrt{- h_5} = L^{-3} r^3 e^{-{ 25 \over 3}\nu } \sqrt{ \f_1 \f_2} $
we get 
\eqn\ccc{
 C_{0123} = L^{-4} r^4  C(r) \ , \ \ \ \ \ \
C(r) =    1  +  \g \bigg[{ 250 \over 32} \rr^8 +
  { 3965 \ov 64}  \rr^{12} \bigg] 
+ O(\g^2) \ .  }
We shall assume that  there are no  direct $\a'$-corrections to the
 form of the D3-brane probe action, i.e.  that 
 only the background  is deformed  by $O(\g)$ terms.  
Then the  static potential part of the action  $I =T^{-1} F$ 
 or the  free energy $F$ is
(cf. \onee)\foot{Note that  in terms of the string-frame 10-d metric
$I= T_3 \int d^4 x \big(  e^{-\p} \sqrt {-  \det_4 G_{10}} - C_{0123} \big) $ but the  dilaton factor  drops out 
 once we transform to the Einstein-frame metric.}
$$F= T_3  V_3 L^{-4} 
r^4 \bigg[ \sqrt { e^{-{ 40 \over 3}\nu(r) } \f (r)  \f_1(r)  } -  C (r)  \bigg]
\   $$
$$
= T_3  V_3 L^{-4} 
r^4 \bigg\{ \sqrt { 1 - \rr^4 } \   \bigg( 1 -  \g \bigg[
   { 75 \ov 2}\rr^4 
   +  { 325\ov 8} \rr^8    - { 155 \ov 8} \rr^{12} \bigg] \bigg) 
$$
\eqn\stat{
 -\  1  -  \g \bigg[ 
{ 125 \over 32}  \rr^8     +   { 3965 \ov 64}   \rr^{12}  \bigg]  
+ O(\g^2) \bigg\} \ .  
}
The leading $O(\g)$   correction  to the strong-coupling
result for the   SYM free energy  in the  `massless' phase 
\nnn\ found in \gkt\ is  (see also \fef,\faf) 
\eqn\old{
   F = -{\te {1 \over 8 \pi^2} } N^2 V_3 L^{-8}  r_0^4 
     (1 + 75 \g ) = -{\te {1 \over 8} } N^2 V_3 T^4 (1 + 15 \g ) \ , 
 }
where we used the  modified relation 
between the temperature and $r_0$ \gkt
\eqn\modi{
 T  = { r_0\ov  \pi L^2} ( 1 + 15 \g)  \ . } 
In the present `separated' case the leading $r$-independent term in the interaction
potential  \stat\ is  found to be (cf. \yey)\foot{Note that 
the non-trivial dependence of  $\nu$  on $r$ 
is irrelevant  to   this order 
as     \raa\  starts with  the $\rr^8$ term.}  
\eqn\resu{
F=  -{\te {1 \over 2 \pi^2} }   T_3 N V_3 L^{-8}  r_0^4   \bigg[ 1 + 75 \g +   O(\g^2) \bigg]   =      -{\te {1 \over 4} }   N  V_3 T^4 \bigg[1 + 15 \g  + O(\g^2)\bigg]   \ . 
}
Up to the coefficient $2N$ instead of $N^2$ (cf.\nnn)
this   is  exactly the same  expression 
as in the  coinciding 3-brane case \old.
This is perfectly consistent with the  expectation
that the functions $f$ and $\td f$ in \feif\ should be the same, 
i.e. that the $U(N+1)$  expression \fef\  for the total 
free energy  should apply  to  both coinciding and separated 
brane cases.

Now we can look  at  the subleading term in the  large distance expansion
of \stat.  We find the following $O(\g)$ correction to \tyte\ (see \modi) 
$$ F_1
=   - {\te {  1 \ov 8}  }
( 1 + {\te { 1425 \ov 4}}   \g + ...)   V_3  T_3 L^{-4}   {  r^8_0 \ov r^4}  
$$
\eqn\tuuy{
= \ -  {\te { 1  \ov 16}} \pi^6
 V_3   N   \l^2  \bigg[ 1 + {\te {  945 \ov 32} } \zeta (3) \l^{-3/2}  + ... \bigg]   { T^8  \ov U^4}     \  .  }
This determines the first two terms
in  the function  $h_1(\l)$ in \tyyt.

Similar analysis can be repeated  in the case of the 
near-horizon  $r\to r_0$ expansion
of the interaction potential discussed in section  4. 
As follows from \stat, the   leading  $r^4_0$ term  in \uuu\ gets 
multiplied by  $1 + { 4265 \ov 64} \g$. 
To find the modification of the subleading terms one has  
first to put the 6-d part of the metric \mett\ in the `isotropic' form 
as in \coop, thus  obtaining  the $O(\g)$ correction to the relation \huh\
between $r \simeq r_0 $ and the  new coordinate $\r$. Then one needs to express 
the  action \stat\ in terms of $\r$. 
Instead of \huh\ we now get
$$
\rho =  \exp \int {  dr \ov   r \sqrt \f } \sqrt \f_2 e^{-{8\ov 3 }\nu} 
=\exp \int {  dr \ov   r \sqrt { 1 - \rr^4} } \bigg( 1  + \four \g [ 
150 \rr^4 +    145 \rr^8 - 575 \rr^{12} ] \bigg)  \ . 
$$
The final effect of expressing the action in terms of $\r$ and expanding in 
powers of $ M\ov T$  can be seen by using  the leading-order relations
\huh,\nam\ in the $O(\g)$ terms in  \stat:
we get $ 1 + c_n \g$ factors for all  $ ({M\ov T})^n$ 
terms  in the  expansion of  the free energy, confirming 
that $F$  has the general structure  suggested   in \eyt,\finc.

\newsec{Acknowledgements}
We are  grateful to  O. Aharony,  N. Itzhaki,   
I. Klebanov  and M. Peskin
for  useful  and stimulating   discussions.
A.T.  acknowledges  the High Energy Theory Group of Rutgers University
for hospitality during the completion of this work.
S.Y. would like to thank the theory group at SLAC for its
hospitality.
The   work of A.T.  is supported in part
by PPARC,   the European
Commission TMR programme grant ERBFMRX-CT96-0045
and  the INTAS grant No.96-538.
The  work of S.Y.  is  supported in part by the US-Israel
 Binational Science Foundation, by GIF --  
  the German-Israeli Foundation for Scientific Research, 
and by the  Israel Science Foundation.


\vfill\eject
\appendix{A}{One-loop $\N=4$ SYM  finite temperature effective potential }

In general, 
the 
1-loop  free energy of a gas of   $N_B$ 
scalars and   $N_F$ spinors  of the same mass  $M$ 
 has the following high temperature ($T\gg M$)  expansion 
  (see e.g. \kapus)
$$F= - V_3 \bigg[ {\pi^2\ov 90} (N_B + {7\ov 8}  N_F) T^4 
-  {1\ov 24}  (N_B + { 1 \ov 2}  N_F)  M^2 T^2   +  {1\ov 12 \pi}   N_B  M^3 T 
$$ 
\eqn\eff{
 +  \ {1\ov 32\pi^2} (N_B-N_F) M^4 \ln  {M\ov \pi T}
 -   {1\ov 16\pi^2} N_B  M^4 \ln 2 + ... \bigg] 
\ . }
The supersymmetric case  corresponds to  $N_B=N_F$.
In the case of $\N=4$   $U(N)$ 
SYM  theory in the vacuum with 
zero scalar values one has   $N_B=N_F= 8 N^2$.
In the  vacuum with the scalar  value  $U$ 
breaking $U(N+1)$ to $U(N) \times  U(1)$ 
 the  contribution of the massive ($M \sim U$) 
  states to the  free energy   is determined by 
\eff\ with  $N_B=N_F= 16 N$.

 In a   general $3+1$ dimensional 
field  theory  with 
 bosonic  and fermionic degrees of freedom  
with   mass operators $\hat M_B$ and $\hat M_F$
(which may depend on  background fields)
the proper-time representation for the  one-loop 
free energy 
has the form
\eqn\field{
 F =  - {V_{3}\ov 2(4\pi)^{2}}
 \int^\infty_0 { d\s \ov \s^{3}}\  \bigg[
 \t_3 ( 0| {i\b^2 \ov 4 \pi \s}) 
  \     \Tr\  e^{- \s \hat M^2_B} 
-
 \t_4 ( 0| {i\b^2 \ov 4 \pi \s}) 
  \     \Tr\  e^{- \s \hat M^2_F}  \bigg]
\ .  }
The inverse temperature $\b = T\inv  $ is the period of the euclidean time direction.
Using the relation  for the  Jacobi $\theta$-functions 
\eqn\ide{
\t_3 (0| iz) \pm  \t_4 (0|iz) =  \sum^\infty_{n=-\infty} [1 \pm (-1)^n]
 e^{- \pi z n^2} =  2 \t_{3,2} (0| 4iz) \ ,  }
in the special case of a free supersymmetric theory with equal 
 bosonic  and fermionic masses $\hat M_B =\hat M_F= \hat M$  this becomes (see e.g. \TT) 
\eqn\eld{
F =   - {V_{3} \ov (4\pi)^{2}}
 \int^\infty_0 { d\s \ov \s^{3}}\  \t_2 ( 0| {i\b^2 \ov  \pi  \s}) 
  \    \Tr\  e^{- \s \hat M^2} 
= - {V_{3} T^4   \ov (4\pi)^{2}}
 \int^\infty_0 { dt \ov t^{3}}\  \t_2 ( 0| {i \ov  \pi t}) 
  \    \Tr\  e^{- t {\hat M^2\ov T^2 } }   \ . 
} 
If the  fields are   massless
 with  total  $N_B$ of   bosonic degrees of freedom 
($N_B=8 N^2$ for   $U(N)$ $\N=4$ SYM) 
\eqn\zzee{
F  = -{\te {  \pi^2 \ov 6}} N_B  V_3  T^4   \ .   }
At  zero temperature  the 1-loop  effective action  of $\N=4$ SYM theory 
in a general  homogeneous 
scalar background has the following form \frt
\eqn\efa{
S_{eff} =  - {V_4 \ov 2 (4\pi)^2} 
\int {d\s \ov \s^3} \  \  \Tr ( e^{-\s \hat M^2_1 }  + 2 e^{- \s\hat  M^2} - \four e^{- \s\hat  M^2_{1/2}} ) \ , 
}
where the mass matrices are
$$
(\hat M^2)^{ab} = f_{aec}f_{bed} A^c_i A^d_i \ , 
\ \ \ \ \ \ \ \ (\hat M^2_{1})^{ab}_{ij}   =\d_{ij} 
 (\hat M^2)^{ab} - 2 F^{ab}_{ij}\ , $$
$$
(\hat M^2_{1/2})_{\a\b}^{ab}  = \delta_{\a\b} 
 (\hat M^2)^{ab} - \ha (\g_{ij})_{\a\b} F^{ab}_{ij}\ .  $$ 
Here $F^{ab}_{ij} = [A_i, A_j]^{ab} = f^{acb} f^{ced} A^e_i A^d_j$, \ \ 
$i=1,..., 6$ are the internal 6-space indices, 
 $a,b, ...$ are the group indices, 
and $\a,\b=1,...,32$ are the 10-d  spinor indices.
Traces are taken in respective spaces, e.g., for the unit operators\ 
$\Tr ( I_1 + 2 I_0 - \four  I_{1/2}) =N^2 ( 6 + 2 - \four\cdot 32)=0$. 
The  expression \efa\ is obtained \frt\ 
by using 
that the  $\N=4$ SYM is a dimensional reduction of the $D=10$ SYM theory.
In particular, the scalar matrix $A^a_i$ is just
the internal component of the 10-d vector potential.
Note that the mass matrix does not contain 
$g_{\rm YM}^2$ factor 
($g_{\rm YM}^{-2}$  appears in front of the whole action). 

In the  case of a   vacuum  (Cartan direction)  scalar  condensate 
(corresponding to separated parallel  D3-brane configuration)     one has 
$F^{ab}_{ij}=0$\  so that  all mass matrices  coincide (reflecting the remaining supersymmetry).
UV divergences cancel as a result of the $\N=4$ sum rules:
$$
\Tr[(\hat  M^2_1)^k   + 2 (\hat M^2)^k  - \four (\hat M^2)^k ] =0 \ , \ \ \ \  \ \ \ k=0,...,3\ . $$
To obtain the finite temperature analogue 
of \efa\  one  is  simply to add the  
thermal $\theta$-function factors  in \field\
inside the integral in \efa\ 
and to use that $V_4= V_3 \b$
\eqn\fild{
F =  - {V_{3}  \ov 2(4\pi)^{2}}
 \int^\infty_0 { d\s \ov \s^{3}}\  \bigg[
 \t_3 ( 0| {i\b^2 \ov 4 \pi \s}) 
  \     \Tr\ ( e^{- \s \hat M^2_1}   + 2  e^{- \s \hat M^2} )
- \four  \t_4 ( 0| {i\b^2 \ov 4 \pi \s}) 
  \     \Tr\  e^{- \s \hat M^2_{1/2}}  \bigg]
\ .  }
Assuming $F_{ij} =0$ and doing the  traces over $SO(6)$ and spinor indices we get (cf.\eld) 
\eqn\fld{
F =  - {8 V_{3}   \ov 2(4\pi)^{2}}
 \int^\infty_0 { d\s \ov \s^{3}}\  
 \t_2 ( 0| {i\b^2 \ov \pi \s}) 
  \     \Tr\   e^{- \s \hat M^2} 
\ ,   }
where the remaining trace is over the  group indices.
We shall consider the case of a simple  
$U(N+1) \to U(N)\times U(1)$ breaking 
(a single  brane separated  from $N$ others) 
 when the matrix  $\hat M^2$ has 
only one non-zero 
  $2N$-degenerate eigenvalue $M^2$.
Then $F$ is given by the sum of the massless and massive 
 contributions
\eqn\fer{
F= F_0 + F_{M} =
  - {8 V_{3}   \ov 2(4\pi)^{2}}
 \int^\infty_0 { d\s \ov \s^{3}}\  
 \  \t_2 ( 0| {i\b^2 \ov \pi \s})  \  \bigg[ (N^2 + 1) +  2 N 
     \   e^{- \s  M^2}  \bigg]
\ .   }
To expand the massive  contribution
 at  large $T\gg M $ 
one  may rescale $\s$ so that 
\eqn\fled{
F_M =  - {16 N V_{3}  T^4   \ov 2(4\pi)^{2}}
 \int^\infty_0 { dt \ov t^{3}}\  
 \t_2 ( 0| {i \ov \pi t}) 
  \       e^{- t  { M^2 \ov T^2}} 
\ ,   }
 getting 
\eqn\fgf{
F_M (T \gg M) \  = \   - {\te{\pi^2 \ov 6} } 2N V_3 \bigg( 
  T^4   -  3\pi^{-2}  T^2 M^2 +  4 \pi^{-3}  T M^3  + ...\bigg)  \ . 
}
For small $T \ll M $ we  use that \ 
$
 \t_2 ( 0| {i\b^2 \ov \pi \s})_{\b \to \infty} 
\   \to\ \  2 e^{- {\pi \b^2\ov 4 \s} }    , 
$
and the  resulting integral  is  proportional to  
$$
\int^\infty_0  d \s\ \s^{-3} \   \exp ( - \s M^2  - {\pi \b^2\ov 4\s}) = 
16 {M^2\ov \pi\b^2} K_2 ( \sqrt \pi \b M ) \ . 
$$
It may be evaluated,   e.g., by the saddle-point method. 
Instead of having power-series expansion in $ {M\ov T}$, \  for $M\gg T$ 
 the free energy  thus 
 goes  to zero  exponentially  as 
$
 V_3 M^2 T^2  \exp ( - \sqrt \pi {M\ov T}) .   
$


\vfill\eject
\listrefs
\end